\documentstyle[epsfig]{mn}

\input{epsf}

\def\lsim{\mathrel{\hbox{\rlap{\hbox{\lower4pt\hbox{$\sim$}}}\hbox{$<$}}}}
\def\gsim{\mathrel{\hbox{\rlap{\hbox{\lower4pt\hbox{$\sim$}}}\hbox{$>$}}}}

\def\and   {\rm {et al.} \rm}  
\def\etal  {\rm {et al.} \rm}

\begin{document}

\title[The Durham/UKST Galaxy Redshift Survey]
{The Durham/UKST Galaxy Redshift Survey - VI.
Power spectrum analysis of clustering}

\author[F. Hoyle  et al ]
{
F. Hoyle, C.M. Baugh, T. Shanks, A. Ratcliffe
\\
Department of Physics, Science Laboratories, South Road, Durham DH1 3LE.
\\
email fiona.hoyle@durham.ac.uk
}
\maketitle 

\begin{abstract}
 We present the power spectrum analysis of clustering in
 the Durham/UKST Galaxy Redshift Survey.
 The Survey covers 1450 square degrees and consists of 2501 
 galaxy redshifts. The galaxies are sampled at a rate of 1 in 3
 down to a magnitude limit of $b_{J}$ $\lsim$ 17 from COSMOS
 scanned UK-Schmidt plates.
 Our measurement of the power spectrum is robust for wavenumbers 
 in the range 0.04 $h \,{\rm Mpc}^{-1} \le k \le 0.6 h\,{\rm Mpc}^{-1}$. 
 The slope of the power spectrum for $k > 0.1 h \,{\rm Mpc}^{-1}$ 
 is close to $ k^{-2}$.
 The fluctuations in the galaxy distribution can be expressed as  
 the {\it rms} variance in the number of galaxies in spheres
 of radius 8 $h^{-1}\,{\rm Mpc}$ as $\sigma_{8} = 1.01 \pm 0.17 $.
 We find remarkably good agreement between the power spectrum measured
 for the Durham/UKST Survey and those obtained from other optical studies on 
 scales up to $\lambda=2\pi/k \sim 80 h^{-1}\,{\rm Mpc}$.
 On scales larger than this we find good agreement with the 
 power measured from the Stromlo-APM Survey (Tadros \& Efstathiou), 
 but find more power than estimated from the Las Campanas Redshift Survey 
 (Lin et al).
 The Durham/UKST Survey power spectrum has a higher amplitude than the power 
 spectrum of IRAS galaxies on large scales, implying a relative bias 
 between optically and infra-red selected samples of $b_{\rm{rel}}=1.3$.
 We apply a simple model for the distortion of the pattern of clustering 
 caused by the peculiar motions of galaxies to the APM Galaxy Survey power 
 spectrum, which is free from such effects, and find a shape and amplitude 
 that is in very good agreement with the power spectrum of the 
 Durham/UKST Survey. 
 This implies $\beta=\Omega^{0.6}/b=0.60\pm0.35$, where $b$ 
 is the bias between 
 fluctuations in the galaxy and mass distributions,  
 and also suggests a one dimensional velocity dispersion of 
 $\sigma = 320 \pm 140 {\rm kms}^{-1}$.
 We compare the Durham/UKST power spectrum with Cold
 Dark Matter models of structure formation, including the effects of nonlinear
 growth of the density fluctuations and redshift-space distortions on the
 theoretical power spectra.
 We find that for any choice of normalisation, the standard CDM model 
 has a shape that cannot be reconciled with the Durham/UKST Survey power 
 spectrum, unless either unacceptably high values of the one dimensional 
 velocity dispersion are adopted or the assumption that bias is 
 constant is invalid on scales greater than $ 20 h^{-1}\,{\rm Mpc}$. 
 Over the range of wavenumbers for which we have a robust measurement 
 of the power spectrum, we find the best agreement is obtained for a 
 critical density CDM model in which the shape of the power spectrum 
 is modified. 
\end{abstract}


\begin{figure*}
\begin{tabular}{cc}
{\epsfxsize=8.truecm \epsfysize=8.truecm 
\epsfbox[70 190 550 600]{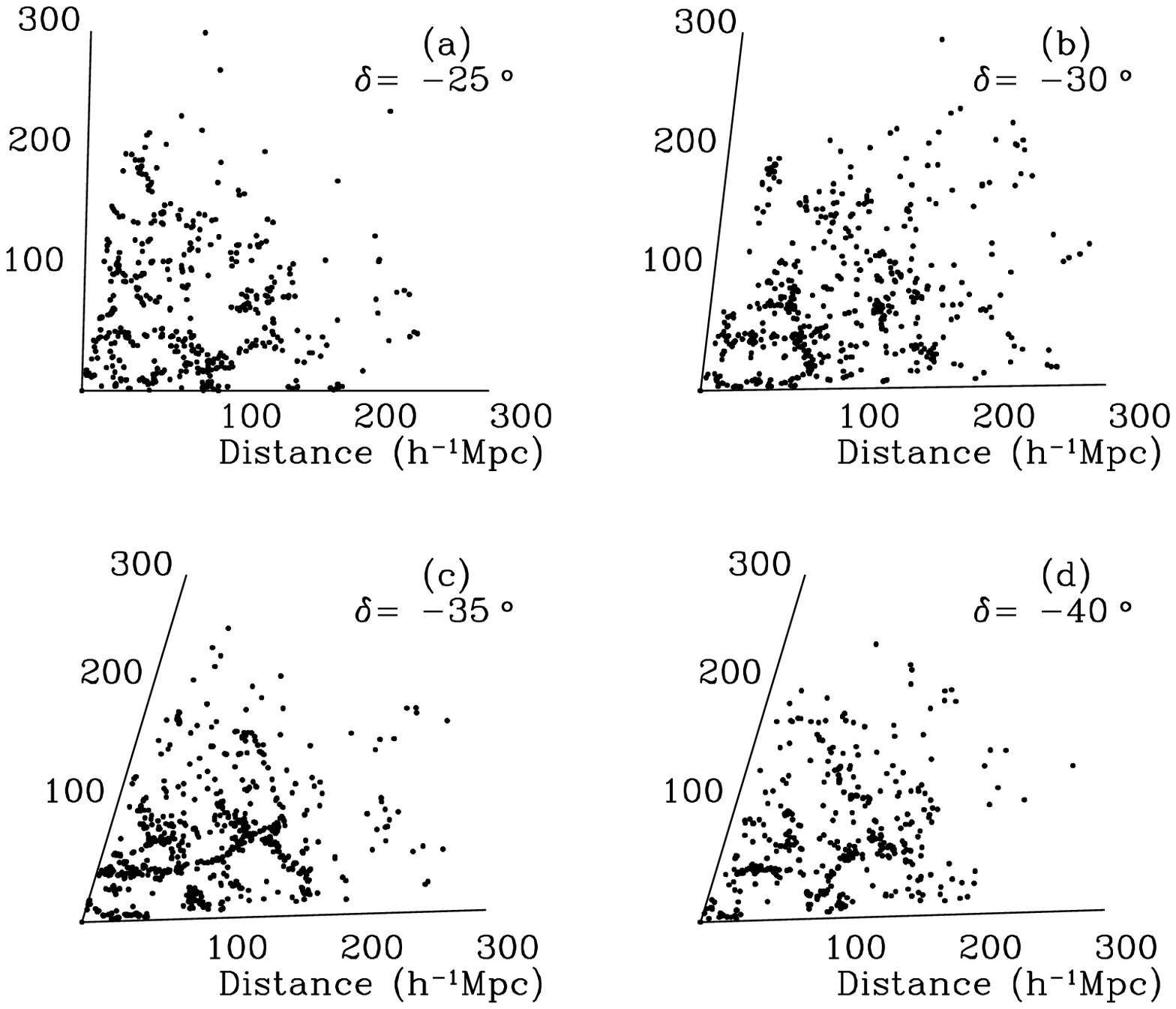}}
&
{\epsfxsize=8.truecm \epsfysize=8.truecm
\epsfbox[70 190 550 600]{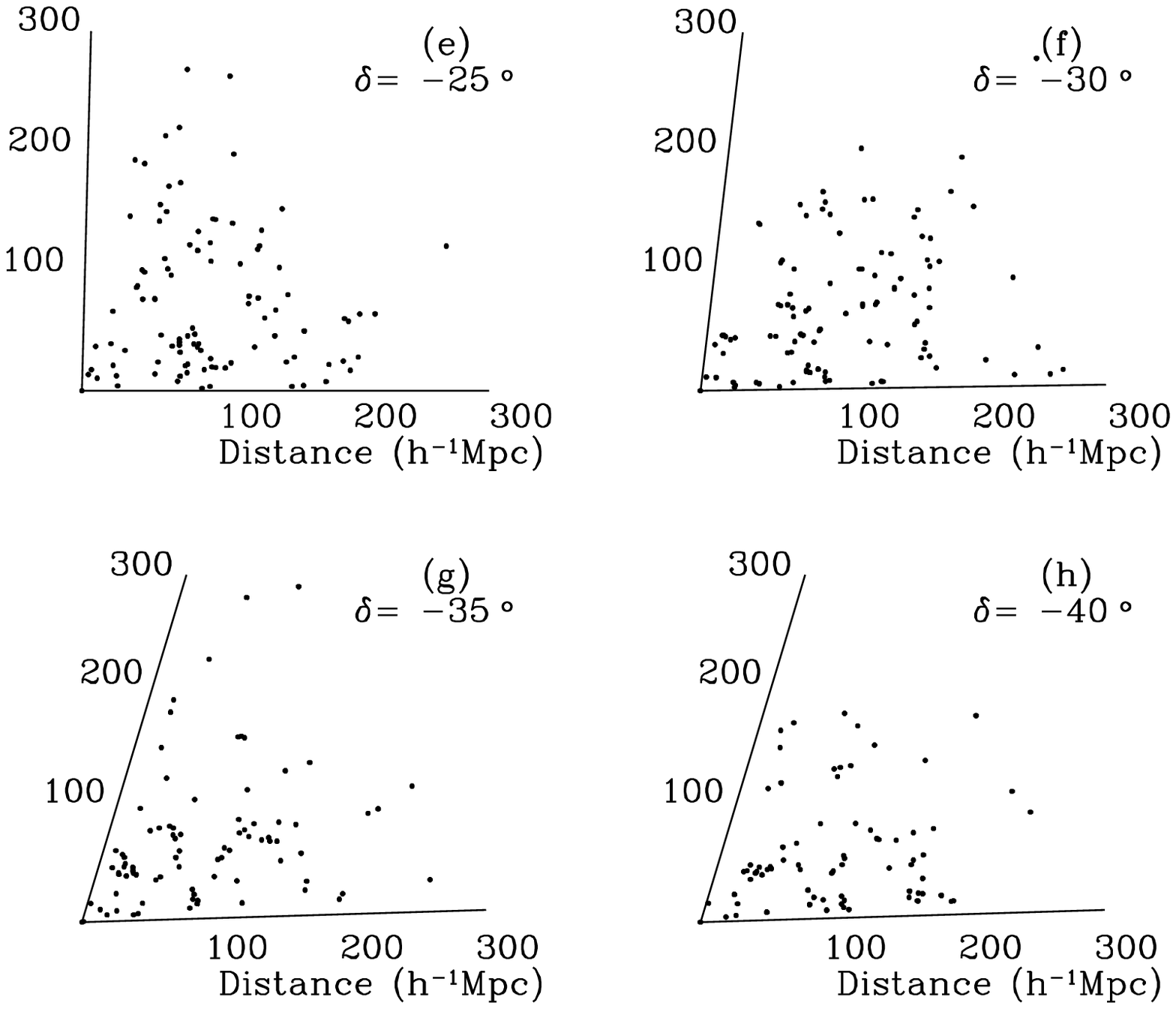}}
\end{tabular}
\caption
{
The four panels (a-d) show the galaxies in the 
Durham/UKST Survey. The four panels (e-h) show 
galaxies in the Stromlo-APM Survey which lie on the same plates.
The declination slices are 5$^{\circ}$ thick and are centred on the
declination shown in each panel. }
\label{fig:gals}
\end{figure*}

\section{Introduction} 
Measuring the primordial power spectrum of density fluctuations 
in the universe is of fundamental importance 
in the development of a model for the formation of large scale structure.
The shape and amplitude of the power spectrum contain information 
about the nature of dark matter and the relative densities 
of dark matter and baryons.
Several obstacles prevent a direct measurement of the primordial
power spectrum from surveys of the local universe.
The gravitational amplification of density fluctuations
leads to a coupling of perturbations on different length scales.
This results in a change in the shape of the power spectrum, except
on large scales where the {\it rms} fluctuations are still less
than unity (e.g. Peacock \& Dodds 1994; Baugh \& Efstathiou 1994b).
Structures are mapped out by galaxies and these may be biased
tracers of the underlying mass distribution (Davis et al 1985).
Furthermore, the relation between fluctuations in the galaxy and
mass distributions could be a function of scale and this needs to
be addressed with a model for galaxy formation (e.g. Benson et al 1999).
The pattern of clustering is also distorted when galaxy positions are inferred
directly from their redshifts. This is due to a contribution to the
observed redshift from the peculiar motion of the galaxy, that arises from
inhomogeneities in the local gravitational field, in addition to the
contribution from the Hubble flow (Kaiser 1987; Peacock \& Dodds 1994).

Measurements of galaxy clustering have improved dramatically in the
last ten years with the completion of several large galaxy surveys.
The infra-red selected QDOT redshift survey (Efstathiou et al 1990;
Saunders et al 1991) and the optical, angular APM Survey (Maddox
et al 1990) 
were the first to demonstrate that there was more power in the
galaxy distribution on large scales than expected from the
standard Cold Dark Matter theory of structure formation.
This led to variants of the standard CDM picture being considered.

The power spectrum has become the favoured statistic for quantifying
galaxy clustering. This is despite the development of
improved estimators for the two-point correlation 
function (Hamilton 1993; Landy \& Szalay 1993).
Both statistics are affected by uncertainties in the mean density of
galaxies, however these uncertainties affect the correlation function
on all scales whereas they only affect the power spectrum on large
scales.
The power spectrum is also the quantity directly predicted by theory.
Errors in the power spectrum are essentially uncorrelated, before 
the mixing of different Fourier modes due to the convolution
of the power spectrum of galaxy clustering with the power
spectrum of the survey window function.
Power spectra are usually estimated by a Fast Fourier transform 
(FFT) and are therefore relatively quick to compute.
Recent theoretical work (Tegmark et al 1998) has demonstrated
that power spectrum analysis can be extended to adjust for various
systematic effects and biases in the data, such as obscuration by dust
or the integral constraint, which we discuss in Section \ref{s:tests}.
However, in general these corrections require an assumption about the
form of the underlying power spectrum and are therefore model dependent.
For this reason, and because the more advanced analysis outlined
by Tegmark et al (1998) has yet to be applied to any existing galaxy survey 
to enable a comparison,
we follow the approach developed by Feldman, Kaiser \& Peacock (1994) and
Tadros \& Efstathiou (1996).

We apply power spectrum analysis to the Durham-/UKST galaxy redshift
survey. The clustering of galaxies in this survey has been studied
using the two point correlation function in earlier papers of this
series (Ratcliffe et al 1996; 1998b);
the magnitude of redshift space distortions was considered in
Ratcliffe et al (1998c).
Although the two point correlation function is the Fourier transform
of the power spectrum, the same is not true of a noisy estimate of
the two-point function.
In addition to studying a flux limited sample, in which the
galaxies are weighted such that the variance in the power spectrum estimate
is minimised, we also study volume limited samples, in which all galaxies
are given equal weight.

In Section 2, we describe the Durham/UKST Survey. 
We outline the construction of different subsamples of the Survey for 
power spectrum analysis in Section 3.
Power spectrum estimators are tested using mock catalogues drawn 
from a large numerical simulation of clustering in Section 4 and we 
present our results in Section 5 and compare with other surveys in Section 6.
The implications for models of large scale structure formation in the
universe are discussed in Section 7 and our
conclusions are given in Section 8.

\section{The Durham/UKST Survey}
\label{s:survey}

\begin{figure}
{\epsfxsize=8.5truecm \epsfysize=8.5truecm 
\epsfbox[70 190 550 600]{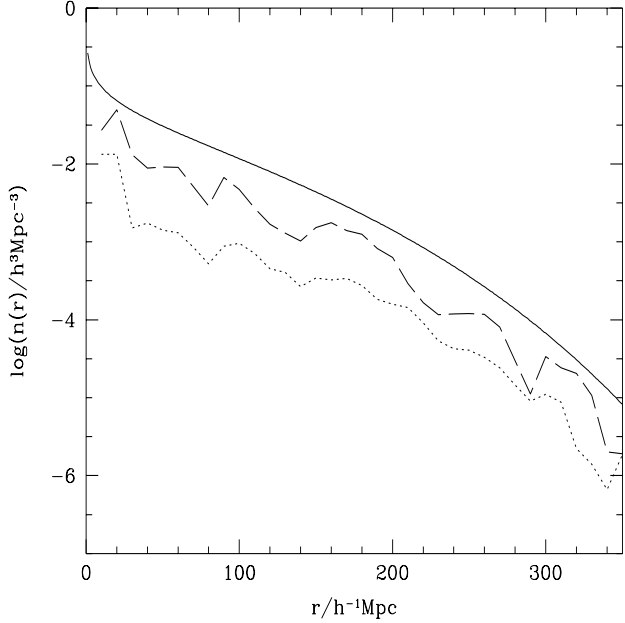}}
\caption
{
The solid line shows the radial number density of 
galaxies to a magnitude limit of $b_{J} \sim 17$, computed 
using the luminosity function of Ratcliffe et al (1998a). 
The dashed line shows the observed number density of Durham/UKST galaxies, 
which  are sampled at a rate of 1 in 3 from the EDSGC catalogue to 
this magnitude limit. 
The dotted line shows the radial number density of galaxies in the 
Stromlo-APM Survey, which are sampled at a rate of 1 in 20 from the 
APM catalogue to approximately the same magnitude limit.
}
\label{fig:numden}
\end{figure}

Full details of the construction of the Durham/UKST Survey, 
including the tests made of the accuracy of the measured redshifts 
and of the galaxy photometry can be found in the earlier
papers of this series (Ratcliffe et al 1996, 1998a,b,c,d; see also 
Ratcliffe 1996).
Here, we restrict ourselves to a summary of
the properties of the survey that are most pertinent to a
power spectrum analysis of galaxy clustering.

The Durham/UKST Survey consists of 2501 galaxy redshifts 
measured with the FLAIR fibre optic system (Parker \& Watson 1995).
The galaxies are sampled at a rate of 1 in 3 down to a magnitude  
limit of $b_{J} \sim 17$ from the parent Edinburgh-Durham Southern 
Galaxy Catalogue (EDSGC;  Collins, Heydon-Dumbleton \& MacGillivray 1988; 
Collins, Nichol \& Lumsden 1992). 
The EDSGC consists of 60 contiguous UK Schmidt Telescope (UKST)  
plates in four declination slices, 
covering a solid angle of $\sim 1450$ square degrees. 

In Figure \ref{fig:gals}, we contrast the visual appearance of the 
Durham/UKST Survey (panels a-d) with that of 
Stromlo-APM Survey galaxies (panels e-h) that lie on the
same UK Schmidt plates (Loveday et al 1996). Structures in the Durham/UKST Survey are
clearly easier to pick out by  eye, due to the six times higher
sampling rate compared with that of the Stromlo-APM Survey.
In the slices centered on  $\delta = -30^{\circ}, -35^{\circ}$
and $-40^{\circ}$, the Sculptor void is visible out to 60\,h$^{-1}$\,Mpc.
(Note that we define Hubble's 
constant as $H_{0}=100$\,h$ \,{\rm kms}^{-1}\,{\rm Mpc}^{-1}$.)
The roof of this feature is seen in the $\delta = -25^{\circ}$ slice. 

The radial number density of galaxies in the EDSGC is shown by 
the solid line in Figure \ref{fig:numden}, which we have computed 
using the luminosity function parameters given by Ratcliffe et al (1998a).
The observed radial number density of galaxies, in bins of 
$\Delta r = 10 \, $h$^{-1}\,$Mpc, is shown by the dashed 
line for the Durham/UKST 
Survey and by the dotted line for the Stromlo-APM Survey. 
The amplitude of the Durham/UKST dashed line lies a 
factor of three below the solid
line due to the sampling rate used. The Stromlo-APM dotted line lies
approximately a factor of 20 below the solid line because the two
surveys have slightly different Schechter function parameters and
magnitude limits.

\section{Power spectrum analysis}
\label{s:psa}

\begin{figure}
{\epsfxsize=8.4truecm \epsfysize=8.4truecm 
\epsfbox[90 190 550 600]{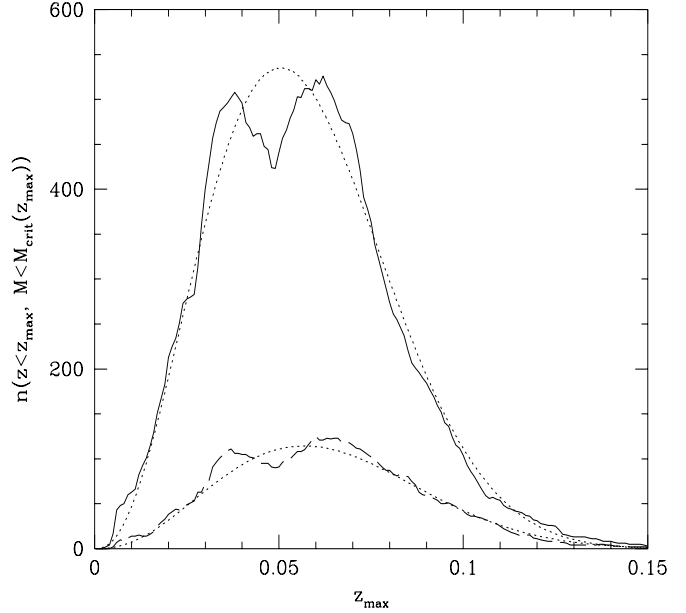}}
\caption
{
The number of galaxies in volume limited samples as a function of the 
redshift used to define the volume limit, $z_{\mathrm{max}}$.
The solid line shows the number of galaxies in volume limited samples 
drawn from the Durham/UKST Survey. 
The dashed line shows the number of galaxies from 
the Stromlo-APM Survey that satisfy the volume limit constraints, 
and which lie on the same Schmidt plates.
The number of galaxies in both cases peaks for a sample limited 
at $z_{\mathrm{max}} = 0.06$ -- there is also a strong feature that can be 
seen in the two catalogues around $z_{\mathrm{max}} = 0.04$. The dotted
lines show the expected number of galaxies obtained by integrating over
the luminosity function, taking into account the different sampling
rates of the two surveys.}
\label{fig:zdist}
\end{figure}

\subsection{Sample definition}
We use two types of galaxy sample in our power spectrum analysis of 
the Durham/UKST survey: (i) flux-limited and (ii) volume limited.
In order to estimate the power spectrum of galaxy clustering in these 
samples, we also need to construct sets of unclustered points with 
the same radial and angular selection; this process is described  
in Section \ref{s:win}.

\subsubsection{Flux-limited sample}
\label{s:flux}

In this case, all galaxies with measured redshifts are used. 
A weight is assigned to each galaxy, to take into account the 
radial selection function of the survey. We adopt the form of the weight 
proposed by Feldman, Kaiser \& Peacock (1994), which minimises 
the variance in the estimate of the power spectrum: 

\begin{equation}
w(r_{i}) = \frac{1}{1 + n(r_{i}) P(k)} \, .
\label{eq:fkp}
\end{equation} 
Here $n(r_{i})$ is the mean galaxy density at the position of the 
$i^{\rm th}$ galaxy. This is calculated by integrating over the 
luminosity function of the survey, taking into account the sampling 
rate.
There is a slight difference in the magnitude limit of each Schmidt
plate in the UKST Survey (see Figure 1 of Ratcliffe et al 1998b), 
so a separate radial weight function is computed for each plate.
Ideally, one should use the true power spectrum in the weight
given by equation \ref{eq:fkp}. However, the results are fairly 
insensitive to the exact choice of power spectrum.
Following the approach taken by Feldman et al (1994) and by Tadros \&
Efstathiou (1996), we adopt a range of constant values of $P(k)$ that are
representative of the amplitude of the power spectrum over
the wavenumbers of interest.
We define the depth of the sample as the distance for which the radial 
weight function $w(r)=0.5$. For our choices of constant power in 
equation \ref{eq:fkp}, this gives depths in the 
range $200$--$320 $\,h$^{-1}$\,Mpc.  
The power spectrum analysis of the flux limited catalogue therefore 
probes volumes in the range $1.2$--$4.9 \times 10^{6} h^{-3} \,{\rm Mpc}^{3}$.

\subsubsection{Volume-limited samples}
\label{s:vol}

The galaxies in a volume limited sample are brighter than 
the apparent magnitude limit of the survey when placed at
any redshift up to that used to set the volume limit, $z \le z_{\mathrm{max}}$.
Hence, as well as requiring that a galaxy have a redshift
$z \le z_{\mathrm{max}}$, the absolute magnitude of the galaxy must
be brighter than:
\begin{equation}
M_{crit} = m_{lim}-25-
5\log_{10}\left[d_l(z_{\mathrm{max}})/h^{-1}{\rm Mpc}\right]- k(z_{\mathrm{max}})  
\end{equation}
where $m_{lim}$ is the magnitude limit of the survey and we use the 
{\it k}-correction given by Ratcliffe et al (1998a) and set $h$=1.
Again, the different plate magnitude limits are taken into
account, so for a given redshift limit $z_{\mathrm{max}}$, the critical
absolute magnitude varies slightly from plate to plate.
We compute the luminosity distance $d_{l}$ assuming an $\Omega_{\circ}=1$ 
cosmology, although our results are insensitive to this choice due to the
relatively low redshift of Durham/UKST galaxies.

In the Durham/UKST Survey, the number of galaxies in a volume limited subset
peaks at a redshift of $z_{\mathrm{max}}=0.06$ (Figure \ref{fig:zdist}).
There are 522 galaxies in this sample.
There is a slightly smaller peak for a sample limited at $z_{\mathrm{max}} = 0.04$.
This feature is particularly strong on the plate centered on
$\delta$=-35$^{\circ}$. 
The same peaks are also seen in volume limited subsamples of the
Stromlo-APM Survey when attention is restricted to those galaxies that lie
on the Schmidt plates covered by the Durham/UKST Survey. The dotted
lines in Fig. \ref{fig:zdist} are theoretical curves calculated by
integrating over the luminosity function. 
The volume limited samples that we consider have maximum 
depths in the range $120$--$230 \,h^{-1}\,$Mpc, and thus sample volumes 
of $0.2$--$1.8 \times 10^{6}h^{-3}\,$Mpc$^{3}$.

\subsection{Survey geometry and radial selection function}
\label{s:win}

The power spectrum measured directly from a galaxy survey
is a convolution of the true power spectrum of galaxy clustering
with the power spectrum of the survey window function.
This is because the Fourier modes are orthogonal within an 
infinite or periodic volume, rather than the complicated
geometry probed by a typical survey. 
The power spectrum of the survey window function is estimated by
placing a large number of unclustered points, typically on the 
order of 100 000,
within the angular area covered by the survey, using the
radial selection function that is appropriate for the galaxy
sample under consideration, as described above.
As before, 
the different magnitude limits of the Schmidt plates in the
Durham/UKST Survey are taken into account when the radial selection
function is calculated.

Fig. \ref{fig:win} shows the power spectrum of the  Durham/UKST 
Survey window function for various volume limited and flux limited samples. 
The top panel shows the power spectra of the window function 
for different volume limited samples. 
The width of the window function power spectrum decreases 
as the volume limit adopted increases.
Figure \ref{fig:win}(b) shows the window function power spectra of 
flux limited samples. 
As the value of the power used in equation \ref{eq:fkp} is 
increased, the flux limited sample has a larger effective depth and 
so the width of the window function is reduced.
There is a relatively small change in the width of the survey 
window function when different samples of the data are considered.
Defining the effective width of the window function as the wavenumber 
at which the power spectrum of the window function falls to half its 
maximum value, we obtain $\delta k \sim 0.015 \,h \,{\rm Mpc}^{-1}$. 
At wavenumber separations smaller than this, our estimates of the 
power will be strongly correlated.
For both flux limited and volume limited samples, the window function 
power spectrum is a very steep power law at wavenumbers 
$k \ge 0.06 \,h \,{\rm Mpc}^{-1}$, varying as $ k^{-4}$.

\subsection{Power spectrum estimation}
\label{s:pk}

\begin{figure}
{\epsfxsize=8.5truecm \epsfysize=8.5truecm 
\epsfbox[70 190 550 600]{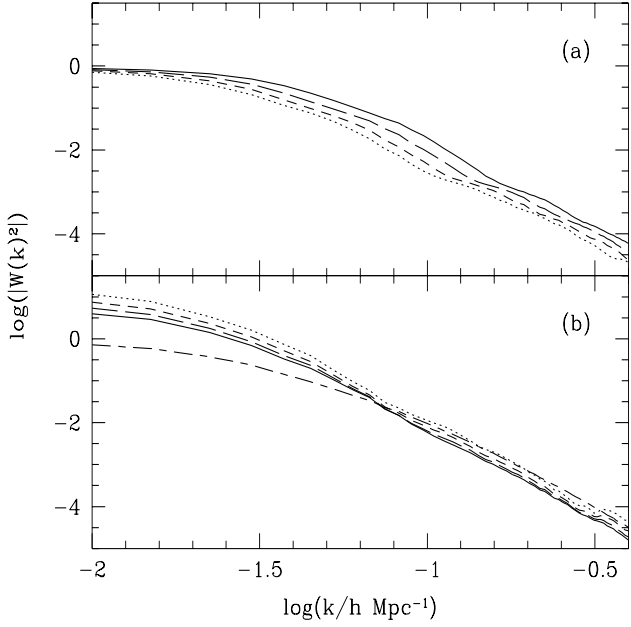}}
\caption
{ The power spectrum of the window function for
different samples extracted from the Durham/UKST survey. 
In (a), the samples are volume limited with a maximum redshift of 
$z_{\mathrm{max}}$=0.05, 0.06, 0.07, 0.08 reading from top
to bottom. 
In (b), we plot the window
function power spectrum for flux limited samples. 
The weights applied are P=32000, 16000, 8000, 
4000 and 0\,$h^{-3}\, {\rm Mpc}^{3}$ 
reading from top to bottom at $\log k = -1.5$. 
For wavenumbers $k \ge 0.06 \,h \,{\rm Mpc}^{-1}$, the window function
power spectrum is a steep power law, $\propto k^{-4}$.}
\label{fig:win}
\end{figure}

The power spectrum estimator that we employ is a generalisation of 
that given by equation 12 of Tadros \& Efstathiou 1996 (see also 
Sutherland \etal 1999), to include the analysis of flux limited 
samples. We do not reproduce the details of their derivation here.

The Fourier transform of the observed galaxy density field, 
within a periodic volume $V$, is given by 
\begin{equation}
\hat{n}_{\circ}({\bf k}) = \frac{1}{V} \sum_i{w_{\rm gal}({\bf x}_{i}) 
e^{\rm i \textbf{k.x$_i$}}}\, ,
\end{equation}
where the weight function $w_{\rm gal}({\bf x}_{i})$ depends upon the type
of galaxy sample under consideration.
For the case of a volume limited sample, $w_{\rm gal}({\bf x}_{i})=1$ for a
galaxy which satisfies the criteria given in Section \ref{s:vol} and
$w_{\rm gal}({\bf x}_{i})=0$ otherwise.
For a flux limited sample, $w_{\rm gal}({\bf x}_{i})$ is given by
equation \ref{eq:fkp}.

The Fourier transform of the survey window function
is approximated by:
\begin{equation}
\hat{W}_e({\bf k}) = \frac{1}{V} \sum_i w_{\rm ran}({\bf x}_{i}) e^{\rm i\textbf{k.x$_i$}},
\end{equation}
where $w_{\rm ran}$ is the weight assigned to one of the unclustered
points used to trace out the survey volume
(note that the definition we have adopted for the Fourier transform of 
the survey window function differs by a factor of $1/\bar{n}_{ran}$ from 
that given in equation 8 of Tadros \& Efstathiou 1996, where 
$\bar{n}_{ran}$ is the number density of unclustered points). 
The power spectra of the survey window function, shown in
Figure \ref{fig:win}, are much steeper than the expected galaxy
power spectrum, falling off as $\propto k^{-4}$ for wavenumbers
$k > 0.06 h \,{\rm Mpc}^{-1}$.
Therefore the main effect of the convolution with the
survey window function is to alter the shape of the power spectrum only
at wavenumbers $k < 0.06 \, h \,{\rm Mpc}^{-1}$.

Following Tadros \& Efstathiou, we define a quantity with a 
mean value of zero:
\begin{equation}
\delta({\bf k})  = \hat{n}_{\circ}({\bf k}) - \alpha \hat{W}_{e}({\bf k}),
\end{equation}
where $\alpha$ is the ratio of the number of galaxies to random points in 
volume limited samples, or the ratio of the sum of the weights, given by 
equation 1, for galaxies and random points in flux limited samples.
The power spectrum of galaxy clustering is then estimated using: 

\begin{eqnarray}
P_{e} (k) &=& \left[  \frac{V}{ S_{\rm ran}^{2} } 
 \sum_{k'} \left( \left| W_{e}(k') \right|^{2} - 
\frac{1}{ S_{\rm ran} } \right) \right]^{-1} 
\nonumber \\
&\times& \left( \frac{ V^{2} |\delta(k)|^{2} }{ S_{\rm gal}^{2} } 
- \frac{1}{ S_{\rm gal} } - \frac{1}{ S_{\rm ran} } \right)
\label{eq:pest}
\end{eqnarray}
where we have used the notation 
$S_{\rm gal} = \sum_{i=1}^{N_{\rm gal}}w^{2}_{\rm gal}$ 
and
$S_{\rm ran} = \sum_{i=1}^{N_{\rm ran}}w^{2}_{\rm ran}$. 
In the case of a volume limited sample $S_{\rm gal} = N_{\rm gal}$, the 
number of galaxies in the sample and 
$S_{\rm ran} = N_{\rm ran}$, the number of unclustered points 
used to define the survey window function. 

The power spectra are computed by embedding the Durham/UKST volume into
a larger cubical volume, $V$.
The density field is typically binned onto a $256^{3}$ mesh
using nearest gridpoint assignment (we discuss the effects of
aliasing and box size in Section \ref{s:tests}).
The Fourier transform is performed with a FFT.

\subsection{Error analysis}
\label{s:err}

\begin{figure}
{\epsfxsize=8.5truecm \epsfysize=8.5truecm 
\epsfbox[70 190 550 600]{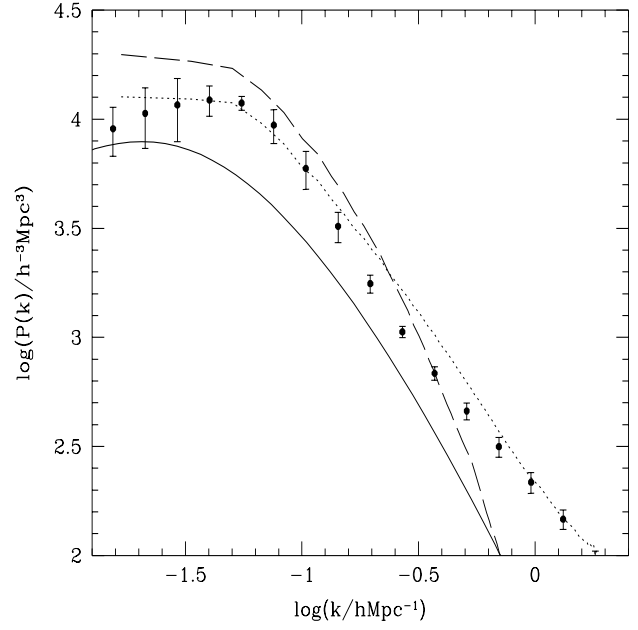}}
\caption
{
The solid line shows the linear power spectrum of the mass
in the Hubble Volume simulation.
The dotted line shows the power spectrum of a subset of the particles 
in the simulation,
selected according to the biasing prescription outlined
in Section \ref{s:err}, measured in a cubical volume of
side $375 h^{-1}\, {\rm Mpc}$.
The dashed line shows the power spectrum of these biased particles 
when the density is binned using redshift space coordinates.
The points show the power spectrum of APM Survey galaxies,
measured in real space. 
  }
\label{fig:bias}
\end{figure}

We estimate the errors on the recovered power spectrum by
constructing mock catalogues that have the same radial and
angular selection as the Durham/UKST Survey and which have approximately the
same clustering amplitude.

We extract mock Durham/UKST catalogues from  
the largest cosmological simulation performed to date, the
{\it Hubble Volume}.\footnote{
The Hubble Volume simulation was performed
by the ``Virgo consortium for cosmological simulations''.
This is an international collaboration involving universities
in the UK, Germany and Canada.
The members of this consortium are: J. Colberg, 
H. Couchman, G. Efstathiou, C. Frenk (PI), A. Jenkins, A. Nelson,
J. Peacock, F. Pearce, P. Thomas, and S. White. G. Evrard is an 
associate member. The Hubble Volume simulation was carried out on the
Cray-T3E at the Max-Planck Rechen Zentrum in Garching.}
The simulation uses $10^{9}$ particles in a
volume of $8 \times 10^{9} h^{-3}\, {\rm Mpc}^3$
and thus contains roughly 10000 independent Durham/UKST Surveys
volume limited to $z_{\mathrm{max}} = 0.06$.
This allows a wide range of clustering environments to be explored,
giving a good assessment of the size of the sampling variance for the
Durham/UKST Survey.

The power spectrum of the Hubble Volume simulation is a variant of
the standard Cold Dark Matter (CDM) model known as $\tau$CDM.
The shape of the power spectrum can be described by the parameter
$\Gamma$, which is set to the value $\Gamma = 0.21$ for $\tau$CDM,
compared with the standard CDM case where $\Gamma = \Omega h = 0.5$ 
(The power spectrum used in the Hubble Volume simulation follows the 
definition of $\Gamma$ used by Efstathiou, Bond \& White 1992).
This change to the power spectrum could be achieved by postulating
a massive neutrino whose decay produces an additional contribution
to the radiation density of the universe, delaying the
epoch of matter radiation equality (White, Gelmini \& Silk 1995).

The {\it rms} density fluctuations in the simulation are set to
be roughly consistent with the local abundance of hot X-ray clusters 
(White, Efstathiou \& Frenk 1993; Eke, Cole \& Frenk 1996). The
variance in the mass contained within spheres of radius $8 h^{-1}\,$Mpc
is $\sigma_{8} = 0.6$. This is smaller than found for the galaxies in
the APM Galaxy Survey, where $\sigma_{8}^{\mathrm{gal}} = 0.84 - 0.96$
(Baugh \& 
Efstathiou 1993; Maddox, Efstathiou \& Sutherland 1996).
In order to make an accurate assessment of the errors in our
recovered power spectrum we need to make mock catalogues in
which the clustering matches as closely as possible that 
in the Durham/UKST Survey.
To extract such catalogues from the Hubble Volume, we apply a
simple biasing prescription to the density field.
We first bin the density field onto a cubical grid of cell size 
$5h^{-1}\,$Mpc, using a nearest gridpoint assignment scheme.
We then associate a probability with each grid cell, which depends on
the ratio of the cell density to the mean density, for selecting a mass
particle from 
that cell to be a biased or `galaxy' particle.
The form of the probability that we adopt is the same as model 1 of
Cole, Hatton, Weinberg \& Frenk (1998) (although these authors apply a Gaussian
filter to smooth the density field - we have chosen the size of our 
cubical grid cell to roughly match the effective 
volume of the Gaussian filter):

\begin{equation}
P(\nu) =  \left\{ \begin{array}{ll}
   \exp(\alpha\nu + \beta\nu^{3/2}) & \mbox{if $\nu\geq0$} \\ 
    \exp(\alpha\nu) & \mbox{otherwise,}
     \end{array} \right. 
\end{equation}
where $\nu$ is the number of standard deviations away from the mean
for the density in the cell and we set $\alpha=1.26$ and $\beta=-0.45$.
The power spectrum of the biased set of particles is shown by the
dotted line in Figure \ref{fig:bias}, which agrees well with the
amplitude of the power spectrum of APM galaxies (Baugh \& Efstathiou 1993;
Gazta\~{n}aga \& Baugh 1998). The dashed line in Figure \ref{fig:bias}
shows the power spectrum of the biased points when redshift space
distortions are also included in the positions of the galaxies. As
expected the power is increased on large scales and damped on small scales.

The errors on the Durham/UKST Survey power spectrum are taken to be the
same size as the fractional errors on the mock catalogue power
spectra. This is a valid assumption when either the contribution of shot
noise to 
the power spectrum is negligible or, as in our case by design, the mock
catalogue power spectrum and the Durham/UKST power spectrum have
similar shapes and amplitudes.
The errors obtained from the mock catalogues converge when averaged over
40 mock observers and are in reasonable agreement with the size of the errors
obtained using the expression given in Equation 2.4.6 of Feldman et al (1994).

\section{Tests of the power spectrum estimation}
\label{s:tests}

\begin{figure}
{\epsfxsize=8.5truecm \epsfysize=8.5truecm 
\epsfbox[70 200 560 600]{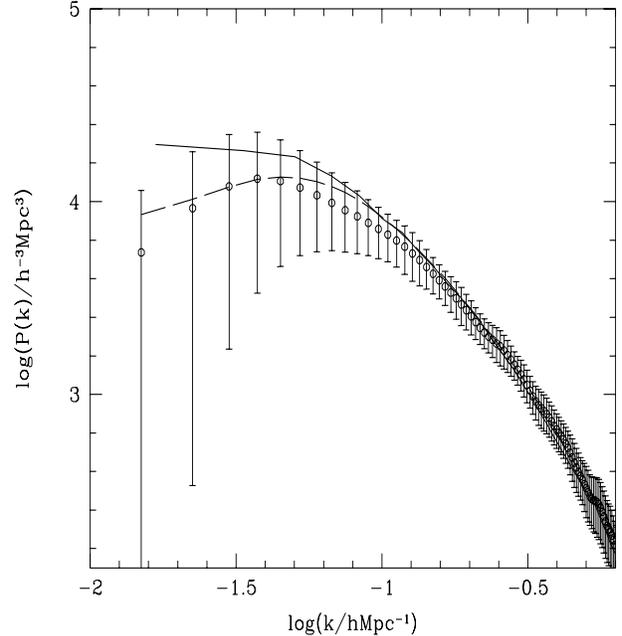}}
\caption
{
The solid line shows the redshift space power spectrum for 
biased particles from the Hubble Volume simulation, 
averaged over 40 cubical volumes of side $375 \, h^{-1}\,$Mpc.
The open circles show the power spectrum averaged over 40 mock Durham/UKST 
catalogues, to a volume limit of $z_{\mathrm{max}}=0.06$. 
The errorbars on these points are the 1$\sigma$ errors for a single power
spectrum extracted from the Durham/UKST survey.
The dashed line shows the convolution of the mean power spectrum measured 
from the large cubical volumes (solid line) with the
window function of the Survey.
}
\label{fig:mock}
\end{figure}

In this Section, we make systematic tests of the power
spectrum estimator (equation \ref{eq:pest}) in order to
assess the range of wavenumbers over which we can make a robust
measurement of the true power spectrum of galaxy clustering.

On large scales there are two main effects that can
cause the recovered power spectrum to differ from the
true power spectrum.
First, Figure \ref{fig:win} shows that the assumption that
the power spectrum of the survey window function is
sharply peaked does not hold for wavenumbers
$k \le 0.04 h \,{\rm Mpc}^{-1}$.
On these scales, the recovered power spectrum
has a different shape to the underlying power spectrum;
the convolution of the power spectrum of the survey window
function with the true galaxy power spectrum alters both the
shape and amplitude of the estimated power spectrum at these
wavenumbers.
Second, the number of galaxies used in
equation \ref{eq:pest} is estimated from the
sample itself. If fluctuations in galaxy density exist on the
scale of the survey, this number can be sensitive to
the environment sampled by the mock catalogue, and hence can
be different from the true mean galaxy density, which is
obtained by considering a much larger volume.
This leads to an underestimate of the power on large
scales (Peacock \& Nicholson 1991; Tadros \& Efstathiou 1996) which is 
sometimes called the integral constraint.
In addition, there will be a contribution to this effect 
from Poisson sampling noise, even in the absence of 
clustering on the scale of the survey.

The redshift space power spectrum of biased tracers of the mass 
distribution in the Hubble Volume simulation is shown 
in Figure \ref{fig:mock}. 
The mean power spectrum is obtained by averaging over 40 cubical 
volumes of side $375h^{-1}\,{\rm Mpc}$.
The power spectrum averaged over 40 mock UKST catalogues made from 
the biased particles is shown by the open circles. 
The errorbars show the $1\sigma$ variance over the 40 mock catalogues.
We have used the number of galaxies in the extracted mock catalogue 
to compute the number density of galaxies for use in the estimator 
(equation \ref{eq:pest}). 
There are still density fluctuations over volumes of the size 
of the Durham/UKST Survey, which leads to a variance in the 
number of galaxies between different mock observers and causes 
a bias in the power spectrum estimate at large scales. 
The dashed line shows the convolution of the mean power spectrum averaged 
over large cubical volumes (shown by the solid line) with the window 
function of the Durham/UKST Survey. 
This shows that the dominant effect on the shape of the power spectrum 
on large scales is the window function convolution rather than the 
integral constraint for the Durham/UKST Survey.
The convolution with the window function power spectrum 
introduces curvature into the recovered power spectrum 
at larger wavenumbers, $ k \sim 0.04 \,h \,{\rm Mpc}^{-1}$,  than the 
real turnover in the $\tau$CDM power spectrum, which 
occurs at $ k \sim 0.02 \, h \,{\rm Mpc}^{-1}$.

\begin{figure}
{\epsfxsize=8.5truecm \epsfysize=9.truecm 
\epsfbox[100 190 550 600]{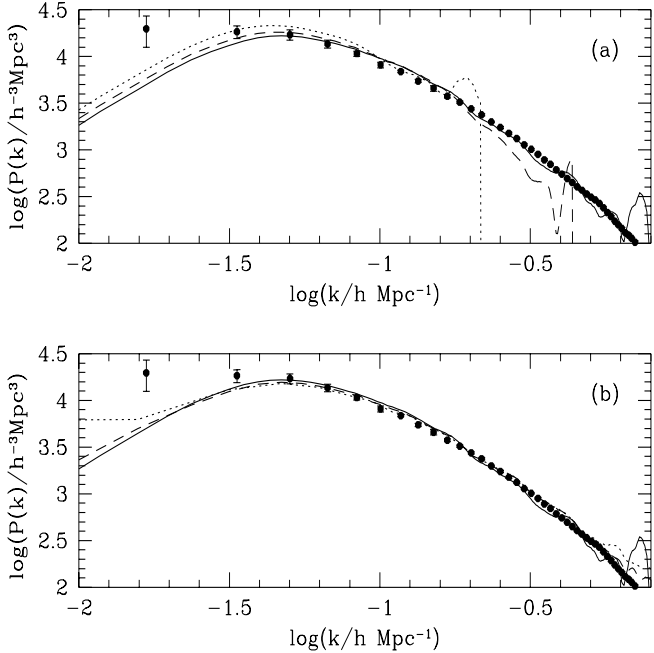}}
\caption
{ 
The points in (a) and (b) show the redshift space power spectrum
of biased particles averaged over 40 large cubical boxes 
extracted from the Hubble Volume simulation. 
(a) shows the effects of changing the size of the FFT grid 
when the mock catalogue is embedded in a fixed size box of side 
$1600h^{-1}\,{\rm Mpc}$ .
The solid line shows the result when the density grid has 
256 cells per side, the dashed line has 128 cells  and the 
dotted line has 64 cells.
(b) shows the effects of varying the size of the transform box 
at a fixed FFT grid size of $256$ cells per side. 
The solid line shows the results for a transform box of 
$1600 h^{-1}\,{\rm Mpc}$, the dashed line for 
$800 \, h^{-1}\,{\rm Mpc}$ and the dotted line for $400 \, h^{-1}\,{\rm Mpc}$.
}
\label{fig:boxsize}
\end{figure}

The Fourier transform of the galaxy density
field is computed by binning the galaxy density field 
onto a finite grid and then performing a Fast Fourier Transform (FFT).
This can lead to spurious features in the power spectrum or 
aliasing of power on scales around the Nyquist frequency of the FFT grid. 
The magnitude of this effect is also sensitive to the scheme used to  
assign galaxies to the density grid.
Figure \ref{fig:boxsize} shows  a series of tests designed to show the 
scales at which aliasing can distort the shape of the recovered 
power spectrum. In both cases we compare the mock catalogue to 
a full box power spectrum which is free from window function effects.
In Figure \ref{fig:boxsize}(a), we vary the dimension of the 
FFT grid within a fixed box size of 1600$h^{-1}\,$Mpc, 
whilst in Figure \ref{fig:boxsize}(b), we vary the size of the 
box in which the mock catalogue is embedded for the 
FFT, and keep the dimension of the FFT grid fixed at $256^3$. 
Figure \ref{fig:boxsize} shows
that using a $256^3$ FFT grid and a box size of $800h^{-1}\,$Mpc, 
gives accurate results down to $k \sim 0.6 h \,{\rm Mpc}^{-1}$ or 
$10 h^{-1}\,$Mpc. 
The power spectra estimated from the mock surveys are discrepant 
with the estimates from large cubical volumes at 
$k \le 0.04 \, h\, {\rm Mpc}^{-1}$ due to the convolution with the 
survey window function.

As we cannot infer the true mean density of galaxies from 
the single observed realisation of the galaxy distribution that we have, 
or equivalently, we do not know the shape of the true power spectrum 
on these scales, we do not attempt to correct the power spectrum at 
large scales for either the `integral constraint' bias or 
for the convolution with the power spectrum of the survey window function. 
Instead, our tests in this section demonstrate that our estimates of the 
power spectrum for the Durham/UKST Survey should be a robust 
measurement of the true galaxy power spectrum over the wavenumber range  
$ 0.04 h \,{\rm Mpc}^{-1} \le \ k \le 0.63 h \,{\rm Mpc}^{-1}$; this 
corresponds to a wavelength range, defined as $ \lambda = 2 \pi/ k$, of 
$160 h^{-1}\,{\rm Mpc}$ to $ 10 h^{-1} \,{\rm Mpc}$; the latter is roughly the 
mean separation of galaxies in a volume limited sample.

\begin{table*}
\begin{tabular}{ccccccc}   \hline
$k h\,{\rm Mpc}^{-1}$ & $P(k)_{vol,z_{\mathrm{max}}0.06} $ & $1\sigma$ & $P(k)_{flux,P=4000} $ & $1\sigma$ & $P(k)_{flux,P=8000} $ & $1\sigma$ \\ \hline
0.0411 &    16153 &    12867 &    24828 &     8946 &    26962 &     9936 \\
0.0561 &    17014 &    10170 &    23596 &     7805 &    24177 &     8556 \\
0.0711 &    13927 &     6467 &    19763 &     6602 &    19775 &     7243 \\
0.0860 &    12215 &     4672 &    15992 &     4880 &    17049 &     5684 \\
0.1000 &    11096 &     3530 &    11332 &     3262 &    12091 &     3606 \\
0.1078 &     9974 &     2962 &     9645 &     2635 &    10103 &     2960 \\
0.1161 &     8682 &     2363 &     7991 &     2011 &     8113 &     2353 \\
0.1251 &     7137 &     1973 &     7446 &     1836 &     7732 &     2190 \\
0.1348 &     5707 &     1571 &     7065 &     1667 &     7481 &     2065 \\
0.1452 &     4546 &     1168 &     6986 &     1515 &     7464 &     2009 \\
0.1565 &     4153 &      934 &     5782 &     1339 &     6012 &     1670 \\
0.1686 &     3937 &      825 &     4477 &     1149 &     4547 &     1351 \\
0.1817 &     3713 &      798 &     3566 &      998 &     3553 &     1150 \\
0.1958 &     3305 &      727 &     3171 &      871 &     3170 &     1051 \\
0.2110 &     2750 &      549 &     2582 &      649 &     2696 &      854 \\
0.2273 &     2189 &      413 &     2039 &      511 &     2230 &      733 \\
0.2449 &     1790 &      371 &     2034 &      530 &     2289 &      779 \\
0.2639 &     1517 &      339 &     1934 &      540 &     2112 &      722 \\
0.2844 &     1247 &      234 &     1572 &      470 &     1727 &      626 \\
0.3064 &     1091 &      184 &     1221 &      371 &     1251 &      456 \\
0.3302 &      932 &      191 &      997 &      372 &     1028 &      474 \\
0.3558 &      907 &      224 &     1262 &      498 &     1451 &      827 \\
0.3834 &      826 &      209 &      882 &      360 &      929 &      514 \\
0.4131 &      745 &      211 &      667 &      251 &      594 &      314 \\
0.4451 &      615 &      213 &      421 &      156 &      344 &      199 \\
0.4796 &      331 &      119 &      311 &      175 &      239 &      237 \\
0.5168 &      181 &       87 &      339 &      170 &      338 &      279 \\
0.5568 &      305 &      196 &      239 &      199 &      206 &      304 \\
0.6000 &      339 &      252 &      111 &      115 &       40 &       89 \\
\hline
\end{tabular}
\caption
{Measurements of the power spectrum from the Durham/UKST Survey.
The first column gives the wavenumber; logarithmically spaced  bins are 
used for $k > 0.1 \, h \, {\rm Mpc}^{-1}$.
The second column gives the power measured in a volume limited 
sample with $z_{\rm max}=0.06$.
The fourth and sixth columns give the power measured in the flux limited 
Durham/UKST Survey, when weights of $P=4000\, h^{-3}\,{\rm Mpc}^{3}$ and 
$P=8000\, h^{-3}{\rm \, Mpc}^{3}$, respectively are used in equation 1.
Columns 3, 5 and 7 gives the $1\sigma$ errors on each measurement.
The errors are the 1-$\sigma$ variance from 40 mock
catalogues extracted from the Hubble Volume. 
}
\label{tab:pk}
\end{table*}

\begin{figure}
{\epsfxsize=8.5truecm \epsfysize=8.5truecm 
\epsfbox[40 190 550 600]{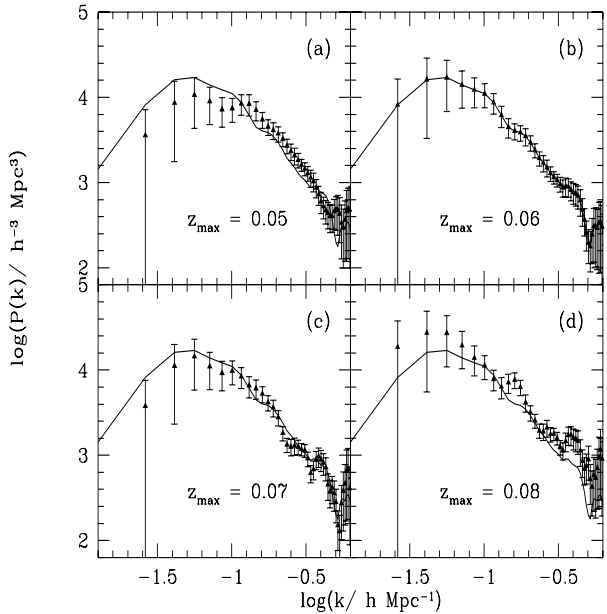}}
\caption
{ 
The power spectrum of the Durham/UKST Survey for 
different volume limited samples. The error bars are the $1 \sigma$ 
variance obtained from the fractional errors on the power found 
in mock catalogues with the same angular and radial selection 
and approximately the same clustering. 
The power spectra are estimated using a box of side $840\, h^{-1}\,$Mpc 
and a $256^{3}$ FFT grid.
The solid line is the mean power for a volume limit defined by 
$z_{\mathrm{max}}=0.06$, the sample that contains the most galaxies, and is 
reproduced in each panel.
}
\label{fig:volps}
\end{figure}

\begin{figure}
{\epsfxsize=8.5truecm \epsfysize=8.5truecm 
\epsfbox[40 190 550 600]{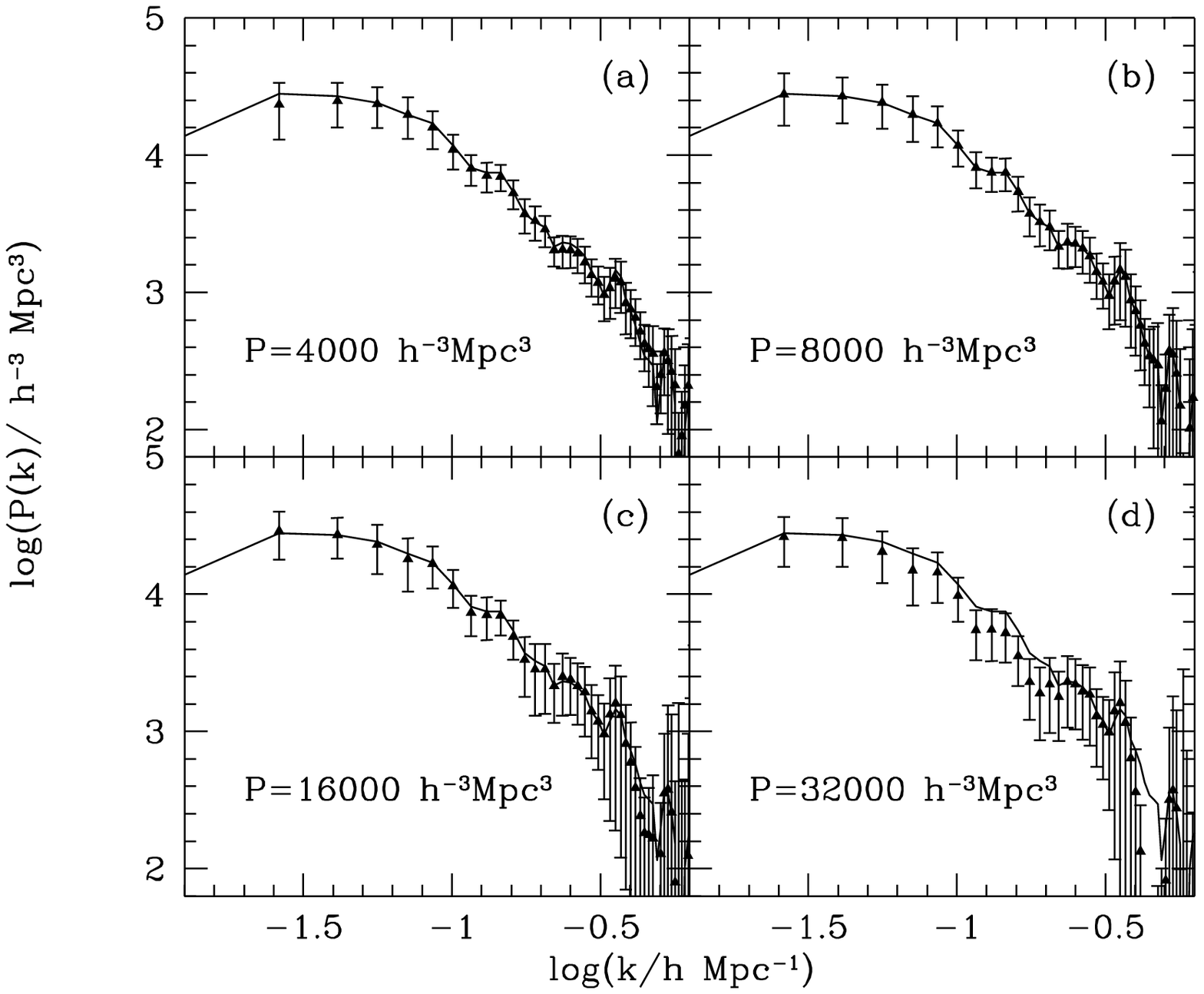}}
\caption
{The power spectrum of the flux limited Durham/ UKST Survey, 
for different constant values of $P(k)$ used in the weight function  
given in equation \ref{eq:fkp}. The values of $P(k)$ used are 4000,
8000, 16000 and 32000 $\,h^{-3}\,$Mpc$^3$ as marked in the panels.
The errorbars show the $1 \sigma$ variance obtained from mock 
catalogues with the same selection and similar clustering.
The solid line is the power spectrum for a weight with 
$P(k) = 8000 \,h^{-3}\,{\rm Mpc}^{3}$ and is reproduced in all 
the panels. The power spectra are computed in a box of side 
$840 \,h^{-1}\,{\rm Mpc}$ using a $256^{3}$ density grid.
}
\label{fig:fluxps}
\end{figure}

\section{Results}
\label{s:res}

In this Section, we analyse volume limited and flux limited samples drawn 
from the Durham/UKST Galaxy Redshift Survey. In all cases, the power 
spectra are computed by embedding the survey in a box of side 
$840 \,h^{-1}\,$Mpc and binning the density of galaxies on a grid of 
256 cells on a side. We have rebinned the estimated power spectrum in 
bins of width $\delta k = 0.015 \,h \,{\rm Mpc}^{-1}$, roughly the width 
at half maximum of the survey window function, in order to reduce 
the correlations between the estimated power at adjacent wavenumbers.
A table of selected results is given in the Appendix.

The power spectra of different volume limited samples of the 
Durham/UKST Survey are shown in Figure \ref{fig:volps}. 
The errorbars are computed using the fractional variance  
in the power averaged over mock catalogues extracted from the Hubble Volume
simulation. 
These mock catalogues were made for each volume limit. 
As discussed in Section \ref{s:err}, these catalogues satisfy the same 
selection criteria and have approximately the same clustering 
as the Durham/UKST Survey galaxies. 
Varying the maximum redshift used to define the volume limited catalogue 
has two effects on the properties of the extracted sample. 
Increasing $z_{\mathrm{max}}$ increases the depth of the sample, 
thereby allowing fluctuations on larger scales to be probed. 
At the same time however, the corresponding absolute magnitude limit 
imposed on the galaxies gets brighter. 
This means that the population of galaxies used to map out the 
clustering varies and it is possible that intrinsically bright 
galaxies could be more strongly clustered than faint galaxies (Park
et al 1994; Loveday et al 1995).
There is a shift in the amplitude of the power spectrum as larger 
values of $z_{\mathrm{max}}$ are considered. However, the power spectra of the 
different samples are all consistent within the $1 \sigma$ errors.

The clustering in the flux limited Durham/UKST Survey is shown in 
Figure \ref{fig:fluxps}. Again, the errorbars show the $1 \sigma$ errors 
obtained from the fractional variance over the power estimated from 
mock catalogues made with the same selection criteria.
The different panels are for weight functions (equation 
\ref{eq:fkp}) using a range of constant values for the power spectrum, 
as indicated in the legend on each panel.
Increasing the value of the power used in the weight, 
causes the weight function to rise at progressively 
larger distances (see Figure 3 of Feldman et al 1994).  
This means that the effective volume probed increases and thus the 
sensitivity to longer wavelength fluctuations increases. 

If there are no systematic problems with the survey, changing the 
value of the power used in the weight function defined by 
equation \ref{eq:fkp} should have little effect upon the 
amplitude of the recovered power spectrum (see the power spectrum analysis 
of the combined 1.2Jy and QDOT surveys by Tadros \& Efstathiou 1995). 
However, the size of the errors on a particular scale will change, depending 
upon whether or not the choice of weight function used 
really is the minimum variance estimator for the amplitude of power at these 
scales. 

The line that is reproduced in each panel of Figure \ref{fig:fluxps} shows the 
power estimated for a weight function with $P(k)=8000 \,h^{-3}\,{\rm Mpc}^{3}$. 
This reference line shows that there is a negligible change in the 
mean power when this weight is varied by a factor of eight over the 
range $P(k)=4000$--$32000 \,h^{-3}\,{\rm Mpc}^{3}$. 
The flux limited power spectrum with a weight $P(k)=4000 \,h^{-3}\,{\rm Mpc}^{3}$ 
has the smallest errorbars over the range of wavenumbers plotted, though 
the errors are not significantly larger for the other estimates of the 
power spectrum. 
The errors on the power spectrum measured from the volume limited 
sample with $z_{\rm max}=0.06$ are larger than the errors on the 
power obtained from the flux limited sample for 
wavenumbers $k < 0.1 \,h \,{\rm Mpc}^{-1}$; however, for 
wavenumbers $k > 0.1 \,h \,{\rm Mpc}^{-1}$ the power spectrum of  
the volume limited sample has smaller errors. 

\begin{figure}
{\epsfxsize=8.5truecm \epsfysize=8.5truecm 
\epsfbox[70 190 550 600]{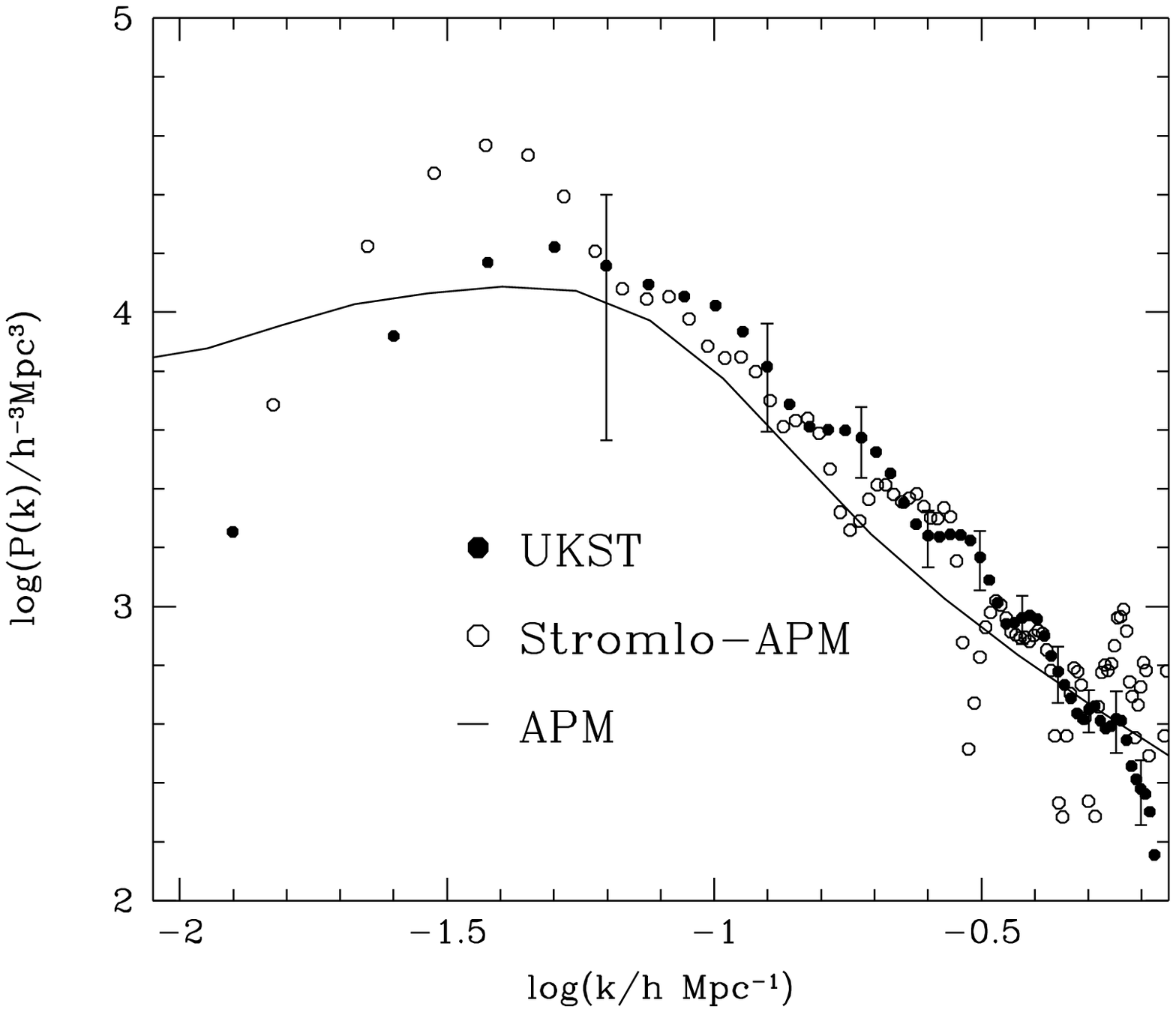}}
\caption
{
The volume limited power spectrum of the
Durham/UKST Survey (solid points) compared 
with the power spectrum of the Stromlo-APM Survey (Tadros
\& Efstathiou 1996); in both cases, the volume limit is
defined by $z_{\rm max} = 0.06$.
The solid line shows the real-space APM galaxy power spectrum
from Baugh \& Efstathiou (1993).
}
\label{fig:su}
\end{figure}

\begin{figure}
{\epsfxsize=8.5truecm \epsfysize=8.5truecm 
\epsfbox[70 190 550 600]{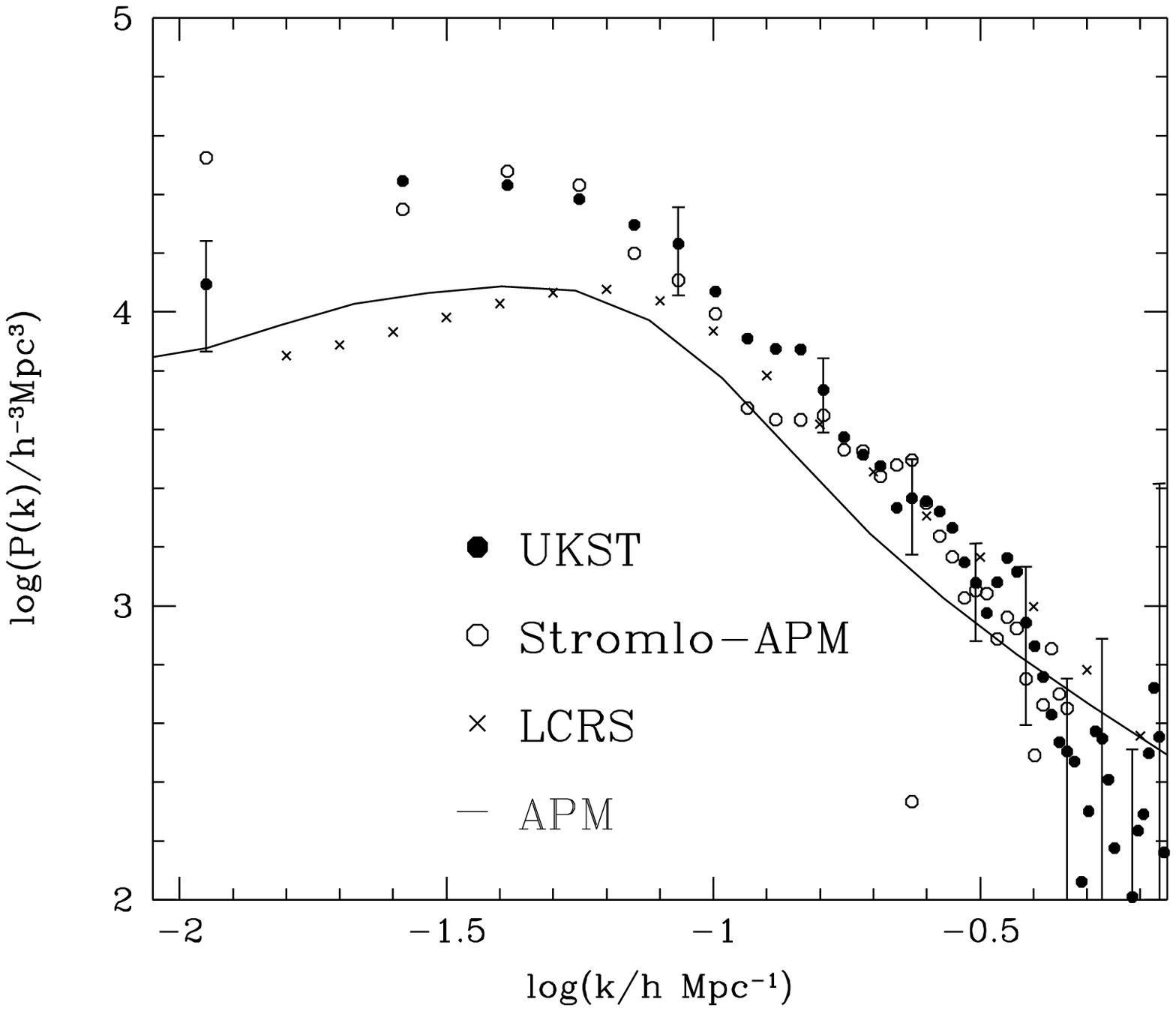}}
\caption
{
The flux limited, $P=8000 \,h^{-3}\,{\rm Mpc}^{3}$, power spectrum of the
Durham/UKST Survey (solid points) compared 
with the flux limited power spectra of other optical samples.
The open circles show the power spectrum of the Stromlo-APM Survey
(Tadros \& Efstathiou 1996), again flux limited with
$P=8000 \,h^{-3}\,{\rm Mpc}^{3}$ and the
crosses show the deconvolved $P(k)$ from the 
Las Campanas Redshift Survey from Lin et al (1996).}
\label{fig:suf}
\end{figure}

The comparison between the power spectra of the flux limited and 
volume limited samples is difficult to interpret. 
Neither the volume nor the way in which the volume is weighted 
can be simply related between the two methods for constructing 
galaxy samples.
Furthermore, volume limited samples select intrinsically brighter 
galaxies as the volume is increased and it is possible that these 
galaxies could have different clustering properties compared 
with fainter galaxies.
Nevertheless, the agreement between the power spectra measured 
from the flux and volume limited samples is very good; if we compare 
the power spectrum from the volume limited sample with $z_{\rm max} = 0.06$,
which contains the most galaxies, 
and the power spectrum with the smallest errors from the flux limited 
survey (i.e. with a value of $P(k)=4000 \,h^{-3}\,{\rm Mpc}^{3}$ 
used in the weight function), then the 
level of agreement is within the $1\sigma$ errors. 
This is a further argument against a significant dependence of clustering 
strength upon intrinsic luminosity within the survey.

\section{Comparison with other measurements of the power spectrum}
\label{sec:comp}

\begin{figure}
{\epsfxsize=8.5truecm \epsfysize=8.5truecm 
\epsfbox[70 190 550 600]{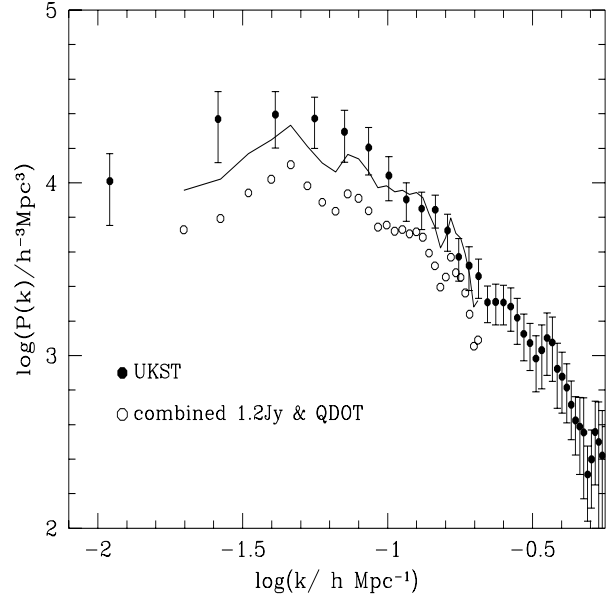}}
\caption
{
The power spectrum of Durham/UKST galaxies (filled circles) compared 
with the power spectrum (open circles) of IRAS galaxies obtained from the  
combined 1.2Jy and QDOT Surveys by Tadros \& Efstathiou (1995). 
Both power spectra are measured from flux limited samples and are minimum 
variance estimates for the respective surveys.
The line shows the IRAS power spectrum after multiplying by a relative 
bias factor squared of $b_{\rm rel}=1.3$, where we have assumed that the 
bias factor is not a function of scale. 
}
\label{fig:qdot}
\end{figure}

We compare our results with measurements of the power spectrum made
from other surveys in Figures \ref{fig:su}, \ref{fig:suf} and \ref{fig:qdot}. 
In Fig. \ref{fig:su}, we compare the power spectrum from a sample of the 
Durham/UKST Survey, defined by a volume limit of $z_{\rm max}=0.06$ 
(filled circles) with the power spectrum of a sample drawn from the 
Stromlo-APM Survey (Tadros \& Efstathiou 1996) with the same 
selection (open circles). The two estimates 
of the power spectrum are in remarkably good agreement, except near 
wavenumbers of $\log k = -0.8, \, -0.5$ and $-0.3$, where there 
are sharp dips in the Stromlo-APM power spectrum.
The solid line shows the real space power spectrum measured from the 
APM Survey (Baugh \& Efstathiou 1993), which is below the power spectra 
measured from the redshift surveys.
We compare estimates of the power spectrum made from flux limited samples 
in Figure \ref{fig:suf}. Again, the filled circles show the power spectrum 
of the Durham/UKST Survey, the open circles show the Stromlo-APM Survey 
and the crosses show the power spectrum measured from the Las Campanas 
Survey (Lin et al 1996). The Durham/UKST and Stromlo-APM Surveys have 
similar magnitude limits, $b_{\rm J} \sim 17$, whereas the Las Campanas 
Survey is approximately $1$--$1.5$ magnitudes deeper,
going to an $R-$band magnitude of $17.3-17.7$, depending
upon the spectrograph used to measure redshifts in a particular field.
The Las Campanas survey consists of six $1.5^{\circ}\times 80^{\circ}$ 
strips and an attempt has been made to deconvolve the survey window 
function to give the estimate of the power spectrum plotted here (Lin et al 1996).
The power spectra from flux limited samples are in good agreement down to 
a wavenumber of $\log k = -1.1$ or for scales $\lambda < 80 \,h^{-1} \,{\rm Mpc}$.
On larger scales than this, the power spectrum measured from the Las 
Campanas Survey is below that obtained from the Durham/UKST and Stromlo-APM 
Surveys, which continue to rise to $\lambda \approx 150 \,h^{-1}\,{\rm Mpc}$. 
On scales larger than this, the convolution with the survey window function 
of these surveys affects the shape of the recovered power spectrum.
Note that the weighting scheme used to estimate the Las Campanas power  
spectrum is different to that employed in this paper, with each galaxy 
weighted by the inverse of the selection function. 

In Figure \ref{fig:qdot}, we compare the power spectrum of 
the Durham/UKST Survey, which is an optically selected sample, 
with the power spectrum obtained from an analysis by Tadros \& 
Efstathiou (1995) of the combined 1.2Jy Survey (Fisher et al 1995) 
and QDOT Survey (Efstathiou et al 1990) datasets, which are selected in 
the infra-red from the IRAS point source catalogue.
We have plotted the minimum variance estimate of the power spectrum 
obtained for each dataset. The filled circles show the Durham/UKST 
power spectrum and the open circles show the power spectrum of IRAS 
galaxies. 
The IRAS galaxy power spectrum has a lower amplitude than the Durham/UKST 
power spectrum. The solid line shows the result of multiplying the 
IRAS power spectrum points by a constant, relative bias factor squared 
of $b_{\rm rel}=1.3$, which agrees with the value inferred by 
Peacock \& Dodds (1994). 

\section{Implications for models of Large Scale Structure}
\label{s:imp}

In this Section we compare the predictions of various scenarios 
for the formation of large scale structure in the universe 
with the power spectrum of the Durham/UKST Survey.

There are several steps that one has to go through in order to 
compare a power spectrum for the mass distribution, calculated 
in linear perturbation theory, with a galaxy power spectrum measured 
using the positions of the galaxies inferred from their redshifts:-

\begin{itemize}
\item[{(i)}] Compute the nonlinear power spectrum of the mass 
distribution given the amplitude of {\it rms} density fluctuations 
specified by the value of $\sigma_{8}$. 
We use the transformation given by Peacock \& Dodds (1996).
\item[{(ii)}] Choose a bias parameter, $b$, relating fluctuations in the 
mass distribution to fluctuations in the galaxy distribution: 
$P_{gal}(k) = b^{2}P_{mass}(k)$. In the following analysis we make 
the simplifying assumption that the bias parameter is independent of 
scale.
\item[{(iii)}] Model the distortion of clustering due to the fact 
that galaxy redshifts have a contribution from motions introduced by
inhomogeneities  
in the local gravitational field of the galaxy as well as from the Hubble flow.
\item[{(iv)}] Convolve the power spectrum with the window function of
the Durham/UKST survey.
\end{itemize}

On large scales, (iii) leads to a boost in the amplitude of the 
power spectrum (Kaiser 1987), whilst on small scales the power is 
damped by random motions inside virialised groups and clusters.
It is important to model these two extremes and the transition 
between them accurately, as this can have a significant effect 
on the shape of the power spectrum over the range of 
scales that we consider.
We model the effects of the peculiar motions of galaxies on the 
measured power spectrum using the formula given by Peacock \& Dodds 
(1994):
\begin{equation}
P_s(k) = b^{2} P_r(k)G(\beta, y) 
\label{eq:biasred}
\end{equation} 
where $P_{s}(k)$ is the galaxy power spectrum measured 
in redshift space and $P_{r}(k)$ is the mass power spectrum 
measured in real space.
The function $G(\beta,y)$, where $\beta = \Omega^{0.6}/b$ 
and $y = k\sigma/100$ ($\sigma$ is the one dimensional 
velocity dispersion), is given by:-
\begin{eqnarray}
G(\beta, y)& = & \frac{\sqrt{\pi}}{8}\frac{{\mathrm{erf}\mathit{(y)}}}{y^5}(3\beta^2 +
4\beta y^2 + 4y^4) \nonumber \\ 
& & - \frac{\exp[-y^2]}{4y^4}(\beta^2(3 + 2y^2) + 4\beta y^2).
\label{eq:redpk}
\end{eqnarray}
This assumes that the small scale peculiar velocities of galaxies are 
independent of separation and have a Gaussian distribution.

We compare the models with the Durham/UKST power spectrum
measured from a sample with a volume limit defined
by $z_{\mathrm{max}}=0.06$.
This power spectrum measurement has larger errors than the minimum
variance power spectrum from the flux limited sample on large
scales, $\lambda=2\pi/k\sim 60\,h^{-1}\,$Mpc. However, on scales smaller
than this, the volume limited power spectrum has the smallest errors.
Furthermore, the fractional errors in the power are smallest  
at high wavenumbers, because these waves are better sampled 
by the survey, and so it is these scales that are the most 
important for constraining the parameters in our model.

\begin{figure}
{\epsfxsize=8.1truecm \epsfysize=8.4truecm 
\epsfbox[90 190 550 600]{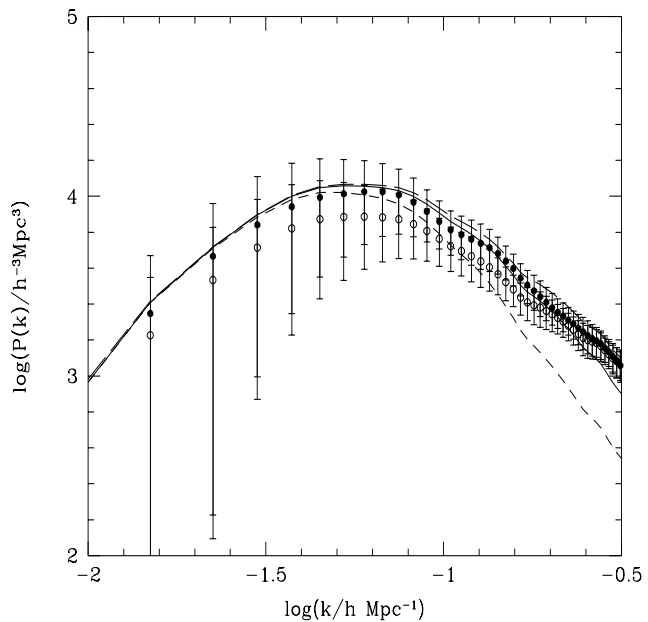}}
\caption
{The open circles show the mean power averaged over 10 mock 
Durham/UKST Surveys in real space and the filled circles are in 
redshift space. The lines show the Peacock and Dodds predictions 
(eqn \ref{eq:biasred}) 
with a bias of $b=1.5$ and $\sigma=300 \, {\rm km\, s}^{1}$ (long dashed), 
$\sigma=500 {\rm \, km\, s}^{-1}$
(solid) and $\sigma=1000 \, {\rm km\, s}^{-1}$ (short dashed).}
\label{fig:conv}
\end{figure}

\begin{figure}
{\epsfxsize=8.4truecm \epsfysize=9.truecm 
\epsfbox[70 190 550 600]{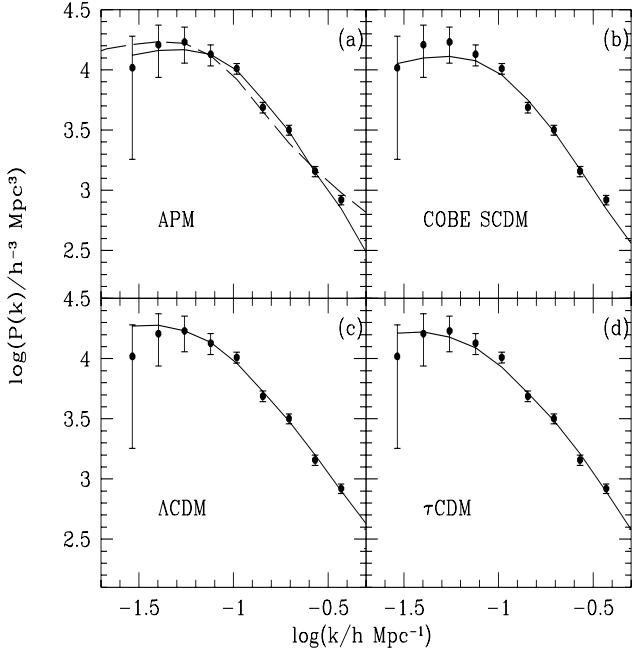}}
\caption
{The points in each panel show the Durham/UKST power spectrum for
a volume limited sample with $z_{\mathrm{max}}$ = 0.06.
The power spectrum estimates have been rebinned to reduce the
covariance in the errors. 
In (a), the dashed line shows the APM galaxy power spectrum
measured in real space, rescaled to match the Durham/UKST power spectrum
at large scales.
The solid line shows the APM power spectrum, including the effects of
distortion in redshift space. 
The remaining panels, b, c, d, show the best fitting curves for several
variants of the Cold Dark Matter model.
Table 1. gives the values of the linear bias $b$
and the one dimensional velocity dispersion $\sigma$ used in
equation \ref {eq:redpk}.}
\label{fig:best}
\end{figure}

We test our simple model for the transformation of a linear theory
power spectrum for mass fluctuations to a galaxy power spectrum
measured in redshift space in Figure \ref{fig:conv}.
The open circles show the mean power spectrum from 10 Durham/UKST
mock catalogues, using the real space coordinates of the particles
to map out the density.
The filled circles show the distortion caused to the power spectrum when
the peculiar motions of the particles are included.
The lines show the results of applying equation \ref{eq:biasred}
to the linear theory $\tau$CDM power spectrum.
This equation results from performing an azimuthal average over the
angle between the line of sight and the wavevector of the density
fluctuation. The observer is also assumed to be at an infinite
distance away from the wave. These two assumptions will mainly affect the
longest wavelength fluctuations in a real survey that does not cover the 
whole sky.
These scales are already distorted by the convolution with the survey
window function.
The model provides a reasonably good fit for a one dimensional
velocity dispersion of $\sigma=500 {\rm kms}^{-1}$, which is
approximately the value found in the simulation (Jenkins et al 1998).

The first test we perform is to compare the power spectrum of 
APM Survey galaxies (Baugh \& Efstathiou 1993, 1994a; Gazta\~{n}aga \& Baugh
1998)  
with the Durham/UKST volume limited power spectrum. The APM power
spectrum is measured in 
real space and is estimated by inverting the angular correlation function 
of APM galaxies with $17 \le b_{J} \le 20$.
The shapes of the real space and redshift space power
spectra can be compared in Figure \ref{fig:best}(a). The
real space power spectrum is shown by the dashed line, after
multiplying by a constant factor of 1.4 to match the Durham/UKST Survey
at small wavenumbers, so that the relative shapes of the real space 
and redshift space power spectra can be readily compared.
We have rebinned the Durham/UKST power spectrum and error bars to match
the binning of the APM power spectrum which has $\delta \log k =
0.13$. 
The spacing of the power spectrum measurements is now much larger 
than the half width of the survey window function, so there is 
essentially no covariance between the errors at different wavenumbers. 
The rebinned Durham/UKST power spectrum is shown in each panel 
of Figure \ref{fig:best} by the points and errorbars. 
We retain this binning of the Durham/UKST power spectrum in the subsequent
analysis of theoretical power spectra below.
As we are comparing two galaxy power spectra, we omit the factor of 
$b^{2}$ in equation \ref{eq:biasred}.
The best fitting APM galaxy power spectrum, including the
redshift space distortions, is shown by the solid line in
Figure \ref{fig:best}(a). The transformation into redshift
space removes the inflection in the real space APM
power spectrum around a wavenumber of $k \sim 0.15 \, h \,{\rm Mpc}^{-1}$.
The best fitting values of $\beta$ and $\sigma$, with $1\sigma$ errors 
are $\beta = 0.60 \pm 0.35 $ and $\sigma=320 \pm 140 {\rm \, km\, s}^{-1}$.
Tadros \& Efstathiou (1996) found $\beta = 0.38 \pm 0.67$ by comparing the 
Stromlo-APM redshift space power spectrum to the APM Survey power spectrum, 
restricting their attention to wavenumbers in the range 
$0.05 < k < 0.1 \, h \,{\rm \, Mpc}^{-1}$, 
on which they argued that the damping 
of power in redshift space is negligible. 
The one dimensional velocity dispersion that we recover from the 
comparison is in excellent agreement with the measurement of 
Ratcliffe et al (1998c), but has much larger errors. 
By considering the galaxy correlation function binned in separation 
parallel and perpendicular to the line of sight, Ratcliffe et al 
obtained a value for the pairwise {\it rms} velocity dispersion along 
the line of sight of $\sigma_{||} = 416 \pm 36 {\rm \, km\, s}^{-1}$. 
This quantity is approximately $\sqrt{2}$ times the one dimensional velocity 
dispersion that we use, giving $\sigma = 294 \pm 25 {\rm \, km\, s}^{-1}$. 
If we add in quadrature the estimated error in the measured redshifts 
$\sim 150 {\rm kms}^{-1}$ (Ratcliffe etal 1998d),
the Ratcliffe et al measurement implies $\sigma = 330 {\rm kms}^{-1}$.

\begin{table*}
\begin{tabular}{llcccccc}  \hline
Model        & $\sigma_8$ & $\Gamma$ & $h$ & $\Omega_{\circ}$ & b              & $\beta$ & $\sigma$ km s$^{-1}$  \\ \hline
SCDM         & 0.52 & 0.5 & 0.50  & 1.0              & 2.97$\pm0.26$  & 0.34 & 1240$\pm200$ \\
COBE-SCDM    & 1.24 & 0.5 & 0.50  & 1.0              & 0.91$\pm0.1$   & 1.10 & 760$\pm130$ \\ 
$\tau$CDM    & 0.52 & 0.2 & 0.50  & 1.0              & 1.64$\pm$0.17  & 0.61 & 320$\pm110$ \\
$\Lambda$CDM & 0.93 & 0.2 & 0.67  & 0.3              & 1.04 $\pm$0.09 & 0.47 & 520$\pm$100 \\
\hline
\label{tab:beta}
\end{tabular}
\caption{The parameters of each of the CDM models and 
the best fitting values of the bias parameter, $b$, and the one
dimensional velocity dispersion, $\sigma$, for various different
cosmological models. The errors are the $1\sigma$ errors obtained using
the errorbars of the Durham/UKST power spectrum. 
We also give the value of $\beta$ implied by our best estimate of the bias 
parameter $b$, for the density parameter $\Omega_{\circ}$ of the model.}
\end{table*}

We also test four popular  Cold Dark Matter (CDM) models 
by treating the bias parameter and the one dimensional
velocity dispersion as free parameters. The mass power spectra 
use the transfer function given in Efstathiou, Bond \& White (1992).
The models that we consider are; $\Omega_{0}=1$ CDM with a
shape parameter $\Gamma=0.5$ and with a normalisation 
of $\sigma_8 = 0.52$ (SCDM) that reproduces the local 
abundance of rich clusters (Eke, Cole \& Frenk 1996); a model with a 
normalisation of $\sigma_{8} = 1.24$ and $\Gamma=0.5$ (COBE-SCDM),
which matches the COBE detection of temperature anisotropies in
the microwave background, but seriously over-predicts the abundance of
hot clusters; $\tau$CDM, with $\Omega_{0}=1$, $\Gamma=0.2$ 
and $\sigma_{8}$ = 0.52, which
simultaneously matches the amplitude implied by COBE and by the
.dat cluster abundance through an adjustment to the shape of the
power spectrum, as described in Section 3.4,
and $\Lambda$CDM, which is a low density model, with a
present day value for the density parameter of $\Omega_{0} = 0.3$ and
a cosmological constant of $\Lambda_{0}/3{\rm H}_{\circ} = 0.7$ (Efstathiou, Sutherland
\& Maddox 1990). The $\Lambda$CDM model has a normalisation of 
$\sigma_{8}=0.93$.

The best fitting parameters are given in Table 1. Note that as we specify 
a value for the density  parameter $\Omega_{0}$ through our choice of 
structure formation model, we are constraining the value of the bias 
parameter $b$; the implied errors on $\beta$ are much smaller than if we 
had not selected a value for $\Omega_{0}$ beforehand.
For all the models considered, reasonable agreement with the 
Durham/UKST Survey power spectrum can be obtained if no restrictions 
are placed on the values of the bias and one dimensional velocity 
dispersion that are used in the fit.
However, the SCDM and COBE-CDM models only produce 
a reasonable fit to the Durham/UKST power spectrum if large 
values of the velocity dispersion are adopted; these values are 
inconsistent with the value we obtain from the comparison with the 
APM Survey power spectrum at more than $3 \sigma$. 
The velocity dispersion required for the $\Lambda$CDM model 
is marginally inconsistent ($1.5 \sigma$) with the value that 
we infer from the comparison with the APM power spectrum.
This agrees with the results of the complementary analysis of the 
two point correlation function carried out by Ratcliffe et al (1998b), 
who analysed the clustering in a N-body simulation with a very similar 
cosmology and power spectrum. 
The $\tau$CDM model gives the best fit to the Durham/UKST data in the 
sense that the values of $\beta$ and $\sigma$ required are in excellent 
agreement with those obtained from the  comparison with the real space galaxy 
power spectrum.

\section{Conclusions}

There is remarkably good agreement between measurements of the power
spectrum of galaxy clustering made from optically selected surveys,
on scales up to $\lambda = 80 h^{-1} \,{\rm Mpc}$.
On larger scales than this, only the most recently completed surveys
cover a large enough volume to permit useful estimates of the
power spectrum to be made.
For scales larger than $\lambda = 80 h^{-1} \,{\rm Mpc}$, we find
good agreement between the power spectra of the Durham/UKST Survey
and of the Stromlo-APM Survey (Tadros \& Efstathiou 1996).
We measure more power on these scales than is found in a clustering
analysis of the Las Campanas Redshift Survey (Lin etal 1996).
We find no convincing evidence for a dependence of galaxy clustering
on intrinsic luminosity within the Durham/UKST Survey. However, we do
measure a higher amplitude for the power spectrum from our
optically selected sample compared with that recovered for
galaxies selected by emission in the infrared; the offset in
amplitude can be described by an optical/infrared bias
factor of $b_{\rm rel}=1.3$.

We have compared the shape and amplitude of the APM Survey power spectrum 
(Baugh \& Efstathiou 1993, 1994; Gazta\~{n}aga \& Baugh 1998), 
which is free from any distortions caused by peculiar velocities, 
with the Durham/UKST power spectrum. The APM power spectrum displays 
an inflection at $k \sim 0.15 \, h \,{\rm Mpc}^{-1}$. Using a simple model 
for the effects of galaxy peculiar velocities that is valid over  
a wide range of scales, we find that the inflection is straightened 
out in redshift space. The APM power spectrum can be distorted to give 
a good match to the Durham/UKST power spectrum for 
$\beta = \Omega^{0.6}/b = 0.60 \pm 0.35$ and a 
one dimensional velocity dispersion of 
$\sigma = 340 \pm 120 {\, \rm km\, s}^{-1}$.
These values are consistent with those found from an independent analysis 
of clustering in the Durham/UKST Survey by Ratcliffe et al (1998c), 
who obtained $\beta = 0.52 \pm 0.39$ (see Hamilton 1998 and references
therein for estimates of $\beta$ made from different surveys using
a range of techniques) and
$v_{12} (\sim \sqrt{2} \sigma) = 416 \pm 36 {\rm \, km \, s}^{-1}$. 
The value of $\beta$ that we obtain from this analysis, can be used, with 
an assumption for the value of $\Omega_{0}$, to infer the amplitude of 
fluctuations in the underlying mass distribution. 
For example, if we assume $\Omega_{0}=1$, our value for 
$\beta$ suggests that APM galaxies are biased with respect to fluctuations 
in the mass by $b = 1.7 \pm 1.0$; this in turn implies a value for the 
{\it rms} fluctuations in mass of $\sigma_{8} = 0.84/b = 0.50 \pm 0.29$, 
which is consistent with that required to reproduce the abundance of 
massive clusters (Eke, Cole \& Frenk 1996). As the abundance of 
clusters and $\beta$ have a similar dependence on $\Omega_{0}$, this 
agreement will hold for any value of $\Omega_{0}$ and therefore does not
constrain $\Omega_{0}$

We have compared theoretical models for structure formation with the 
power spectrum of the Durham/UKST survey. The best agreement is 
found with a variant of the Cold Dark Matter model known as 
$\tau$CDM. 
A low density model with a cosmological constant also provides 
reasonable agreement, but for a velocity dispersion that is marginally 
inconsistent with that obtained from our comparison between the power 
spectra of the Durham/UKST and APM Surveys.
Critical density CDM models with shape parameter $\Gamma=0.5$ require 
one dimensional velocity dispersions that are much too high in order  
to provide a good fit to the Durham/UKST power spectrum.
One possible way to resolve this problem would be to relax the assumption 
that the bias parameter between galaxies and the mass distribution is 
independent of scale. 
Whilst a constant bias is undoubtably a poor approximation on scales 
around a few megaparsecs and smaller (see for example Benson et al 1999), 
our analysis probes scales greater than $20 \, h^{-1}\,{\rm Mpc}$. 
A scale dependent bias on such large scales could be motivated 
in a cooperative galaxy formation picture (Bower, Coles, Frenk \& White 1993), 
though the higher order moments of the galaxy distribution expected in 
such a model are not favoured by current measurements (Frieman \& 
Gazta\~{n}aga 1994).

\section*{Acknowledgments}
We thank the referee, Will Sutherland, for a careful reading of the manuscript 
and helpful comments which have improved the final version of the paper.
FH acknowledges the receipt of a PPARC studentship. 
We would like to thank Helen Tadros for many helpful conversations 
and communicating data in electronic form for the Stromlo-APM
Survey, Huan Lin for supplying LCRS $P(k)$ data, Shaun Cole for
advice and comments on the manuscript and Steve Hatton
for assistance with the error analysis.
Adrian Jenkins made the Hubble Volume simulation available to us,
on behalf of the Virgo Consortium, and kindly provided us with
essential software for reading the simulation output.
We acknowledge the efforts of Alison Broadbent, Anthony Oates, Quentin 
Parker, Fred Watson, Richard Fong and Chris Collins in the construction
of the Durham/UKST Survey.
CMB acknowledges receipt of a computer equipment grant from the
University of Durham.

\end{document}